\documentclass[epsf,psfig,referee]{mn2e}
\usepackage{amssymb,amsmath} 
\usepackage{graphicx}
\usepackage{deluxetable}

\newcommand{\mjup}{$M_{\rm Jup}$}
\newcommand{\mearth}{$M_{\rm Earth}$}

\newcommand{\msini}{$M \sin i~$}

\newcommand{\logrhk}{log $R'_{\rm HK}$}

\newcommand{\p}{258.19$\pm$0.07}
\newcommand{\m}{1.59$\pm$0.02}
\newcommand{\e}{0.233$\pm$0.002}
\newcommand{\om}{352.7$\pm$0.5}
\newcommand{\pt}{10071.0$\pm$0.8}
\newcommand{\sa}{49.5$\pm$0.2}
\newcommand{\au}{0.81$\pm$0.02}

\newcommand{\off}{-3.8$\pm$3.2}
\newcommand{\aoff}{-14.6$\pm$3.2}

\newcommand{\ppe}{5000$\pm$400}
\newcommand{\mme}{0.82$\pm$0.03}
\newcommand{\eee}{0.12$\pm$0.02}

\newcommand{\omome}{195$\pm$48}
\newcommand{\ptpte}{11100$\pm$600}
\newcommand{\sasae}{9.3$\pm$0.3}
\newcommand{\auaue}{5.8$\pm$0.5}

\begin{document}

\title[A long-period planet orbiting a nearby Sun-like star]
{A long-period planet orbiting a nearby Sun-like star}

%Wright?Marcy?

\author[H. R. A. Jones et al.]{
\parbox[t]{\textwidth}{
Hugh R.A. Jones$^1$, R. Paul Butler$^2$,  
C.G. Tinney$^3$, Simon O'Toole$^{1,4}$, Rob Wittenmyer$^3$, Gregory W. Henry$^5$, Stefano Meschiari$^6$, Steve Vogt$^6$, Eugenio Rivera$^6$, Greg Laughlin$^6$,   
Brad D. Carter$^7$, Jeremy Bailey$^3$, James S. Jenkins$^{1,8}$} \\\\
%\vspace*{6pt} 
$^1$Centre for Astrophysics Research, University of Hertfordshire, 
College Lane, Hatfield, Herts AL10 9AB, UK\\
$^2$Carnegie Institution of Washington, Department of 
Terrestrial Magnetism, 5241 Broad Branch Rd NW, Washington, 
DC 20015-1305, USA \\
$^3$Department of Astrophysics, School of Physics, University of NSW, 2052, Australia\\  
$^4$Anglo-Australian Observatory, PO Box 296, Epping 1710, Australia \\ 
$^5$ Center of Excellence in Information Systems, Tennessee State University, 3500 John A. Merritt Blvd., Box 9501, Nashville, TN 37209\\
$^6$UCO/Lick Observatory, Department of Astronomy and Astrophysics, University of California at Santa Cruz, Santa Cruz, CA 95064, USA\\  
$^7$Faculty of Sciences,  University of Southern Queensland, Toowoomba, 
QLD 4350, Australia\\
$^8$Department of Astronomy, Universidad de Chile, Casilla Postal 36D, Santiago, Chile\\
}
\date{}
%$^6$Department of Astronomy, University of California, Berkeley, CA, 94720, USA\\

\maketitle

%add collier-cameron rotation if seems approprate

%\label{firstpage}

\begin{abstract}
The Doppler wobble induced by the extra-solar planet HD 134987b
was first detected by data from the Keck Telescope nearly a decade
ago, and was subsequently confirmed by data from the
Anglo-Australian Telescope. However, as more data have been acquired
for this
star over the years since, the quality of a single Keplerian fit to that
data
has been getting steadily worse. The best fit single Keplerian to the
138 Keck and AAT observations now in hand has an root-mean-square
(RMS) scatter of 6.6 m/s.
This is significantly in excess of both the instrumental precision 
achieved by both
the Keck and Anglo-Australian Planet Searches for stars of this
magnitude, and of the
jitter expected for a star with the properties of HD134987.
However, a double Keplerian (i.e. dual planet) fit delivers a significantly
reduced RMS of 3.3 m/s.
The best-fit double planet solution has minimum planet masses of 1.59 and
0.82~{\m}\mjup, orbital periods of 258 and 5000~d, and eccentricities of 0.23
and 0.12 respectively. We find evidence that activity-induced jitter is a significant factor in our fits and do not find evidence for asteroseismological p-modes. 
We also present seven years of photometry at a typical precision of 0.003~mag with the T8 0.8~m 
automatic photometric telescope at Fairborn observatory.  These observations 
do not detect photometric variability and support the inference that the 
detected radial-velocity periods are due to planetary mass companions rather than due to photospheric spots and plages.

\end{abstract}


\section{Introduction}
Of the more than 300 nearby stars known to harbor one or more planets, 38 are multiple-planet systems, and some show radial velocity residuals indicative 
of additional companions (e.g., Wright et al. 2007). A number of known extrasolar planets (hereafter, shortened to exoplanets) are becoming suitable for increasingly powerful follow-up techniques. Primarily, this has been led by the  discovery of transiting objects though other observations are becoming rewarding. Notably, observations of short-period exoplanets have enabled atmospheric abundance studies (e.g., Swain et al. 2009), phase variation in
flux to be distinguished (e.g., Knutson et al. 2007), and high-resolution
spectroscopic observations may soon allow for direct spectroscopic
detection (e.g. Barnes et al. 2008). 
Recent leaps in image processing (e.g., Marois et al. 2008) have led to companion detections at 0.5 arcsec separations with contrast ratios of 10$^{-6}$ across the near infrared. This has allowed for the direct imaging  of putative exoplanets which in turn is driving the instrumentation programmes of telescopes worldwide.

The architecture of our Solar System is dominated by Jupiter at 5.2~au and by Saturn at 9.5~au. However, very little is known about the frequency or nature of other planetary systems with orbital distances 
greater than 5~au (e.g., Marcy et al. 2005b). Precise radial velocities have only recently reached beyond the 10 years required to sense such objects. For planets orbiting solar mass 
stars, it is key to have sensitivity to Jupiter-like (12~yr) and Saturn-like (30~yr) orbits. 
The ability to put
constraints on such planets, even with potentially incomplete orbits,
will enable imaging observations to target systems for which they can
provide critical observational constraints. With or without imaging, such orbits may substantially constrain the orbital structure and dynamical configuration, providing clues about planetary formation and migration (e.g., Currie 2009) and will help us trace the uniqueness of our own Solar
System. 

The Anglo-Australian Planet Search and the Keck planet searches are long-term radial velocity
projects engaged in the detection and measurement of exoplanets  at the highest possible precisions.
Using the iodine calibration technique (e.g., Butler et al. 1996) they provide coverage of bright inactive F, G, K and M dwarfs.
The AAPS began operation in 1998 January, and is currently surveying 
250 stars. It has published exoplanets with $M$~sin~$i$ ranging from 5\mearth to 10 \mjup 
(Tinney et al. 2001, 2002a, 2002b, 2003, 2005, 2006; Butler et al. 2001, 2002b, 2006a; Jones et al. 2002, 2003a,b, 
2006; Carter et al. 2003; McCarthy et al. 2005; OÕToole et al. 2007, 2009a, 2009b, Bailey et al. 2009, Vogt et al. 2009). With its somewhat longer baseline (since 1996) the Keck project originally announced the exoplanetary signal around HD~134987 (Vogt et al. 2000).  Both Keck and the AAT have been regularly observing the star since 1998.  Nearly a decade later we 
present results for HD~134987 from the combined AAT and Keck dataset.  

\section{Characteristics of HD~134987}
HD134987 (23 Lib) is a solar-type (G4V) star which is nearby (22pc)
and bright (V=6.45), with
low activity (adopted log RHK =--5.1) and high metallicity (adopted
0.25 dex).
As an analogue of the prototype ``Hot Jupiter'' host star (51 Peg), it has long been a target for precision exoplanet surveys. 
Vogt et al. (2000) reported a planetary mass signal with a period $P$~=~259~d, eccentricity $e$~=~0.24, $M$ sin $i$~=~1.58 \mjup ~from Keck data with an RMS of 3~m/s. Butler et al. (2001) conÞrmed the orbit using AAT data. In Butler et al. (2006a) these parameters were revised to  $P$=258 d, eccentricity, $e$~=~0.24, $M$ sin $i$~=~1.64 \mjup\ with an RMS of 4 m/s to a {\bf reduced} $\chi_{\nu}^2$=0.89 fit, including a trend of 2.9$\pm0.2$~m/s per year. At that time the jitter of HD~134987 was estimated to be 3.5m/s. Wright et al. (2007) identify a number of objects whose False Alarm Probability for an additional Keplerian versus a 
simple trend is below 2\%. Wright et al. report that the signal appeared as a change in the level of the residuals between 2000 
and 2002 of 15 m/s. They thus suggested an outer planet on a rather eccentric orbit 
which reached periastron in 2001.  Recently estimates for the jitter, based on Wright (2005),
have been reduced by around a factor of $\sqrt{2}$, (J.Wright,  private
communication)
and an additional 43 epochs have been acquired with the AAT and Keck.

The properties of HD~134987 are summarised in Table
\ref{stellarp}. The variety of recent measurements reflect its inclusion 
in large-scale studies of nearby solar type stars. Recently, Holmberg et al. (2007)
included it in a magnitude-limited, kinematically unbiased study of
16682 nearby F and G dwarf stars and Takeda et al. (2007a) included it 
in a study of the stellar properties for 1040 F, G and K stars observed
for the Anglo-Australian, Lick and Keck planet search programmes. Takeda et al. used high signal-to-noise
echelle spectra (originally taken as iodine-free templates for radial
velocities)
to derive effective temperatures,
surface gravities and metallicities whereas Holmberg et al. 
used Str\"{o}mgren photometry and the infrared flux method calibration of 
Alonso et al. (1996). Both studies used Hipparcos parallaxes to convert 
luminosities in order to make comparisons with different theoretical isochrones and so derive stellar parameters. To determine, stellar masses and ages 
Takeda et al. (2007a), use Yonsei-Yale isochrones (Demarque et al. 2004) and 
Holmberg et al. (2007) use Padova isochrones (Giradi et al. 2000; 
Salasnich et al. 2000). Both sets of derived parameters agree to within
the uncertainties. 
%Holmberg et al. (2008) has reassessed values but a) not available in simbad and b) says it does not change Fe/H so pointless to reference

The low activity index of HD~134987 (\logrhk~$\sim$~--5.1) is consistent with the lack of  significant photometric
variability in measurements made by the Hipparcos satellite.
Combining Hipparcos astrometry with their radial velocities, Holmberg
et al. (2007) and Takeda et al. (2007b) determine U, V, W space velocities consistent with the old disk lifetimes in the range 8-11 Gyr inferred from the isochrones and the lack of X-ray flux detected from HD~134987 (Kashyap, Drake \& Saar 2008).

\section{Spectroscopic Observations}

The 63 epochs of Doppler data obtained at the AAT between 1998
August and
2009 October are shown in Table \ref{aat_vel}. The 75 epochs of
Doppler measurements obtained at
the Keck Telescope between 1996 August and 2009 July are shown in Table \ref{keck_vel}. 
The observing and data processing procedures follow those described by 
Butler et al. (1996, 2001, 2006a).  
All these data have been reprocessed through our frequently up-graded analysis system, and here we report results from the current version of our pipeline.
Our velocity measurements are derived 
by breaking the spectra into several hundred 2~\AA ~chunks and deriving
relative velocities for each chunk. The velocity uncertainty, given in the third 
column and labelled `Unc.'
is determined from the scatter of these chunks.
This uncertainty includes the effects of photon-counting uncertainties,
residual errors in the spectrograph PSF model, and variation in
the underlying spectrum between the template and iodine epochs. For both the AAT and Keck, 
observations in which the uncertainty is more than three-times the median
uncertainty of the entire set are not reported. All
velocities are measured relative to the zero-point defined by
the template observation. Since the AAT and Keck data were processed with different templates,
we treat the difference between
their zero-points as a free parameter in our fitting procedures.

\section{Orbital Solution for HD~134987}

\begin{figure}
%\hspace*{-1cm}
\includegraphics[width=110mm,angle=90]{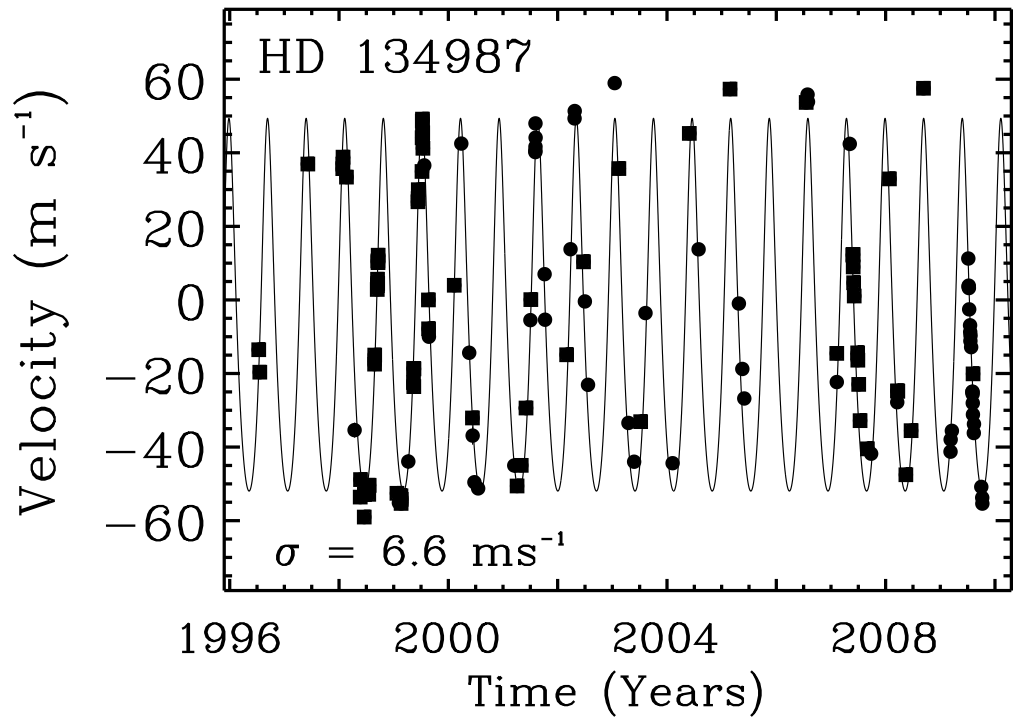}
 \caption{The solid line indicates the best fit single Keplerian which has 
an RMS of 6.6 m/s fit to the data. 
 % with the parameters shown in table 
The Keck data are shown as squares and the AAT data as circles. }
\label{hd134987}
\end{figure}

\begin{figure}
\includegraphics[width=110mm,angle=90]{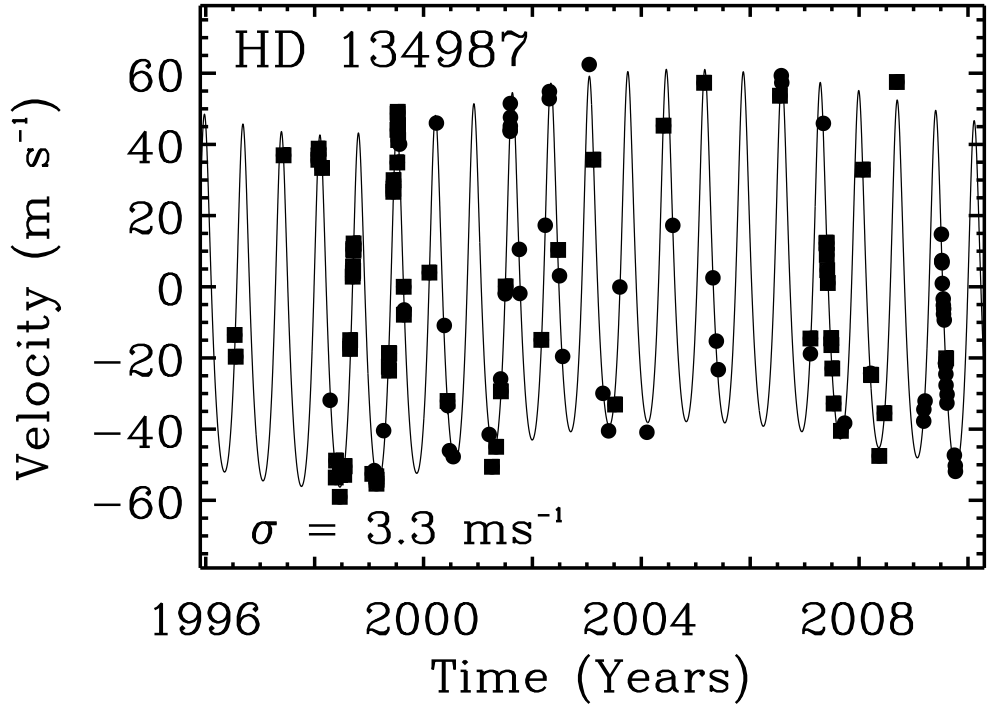}
 \caption{The solid line shows the best fit double Keplerian
to the Keck (squares) and AAT (circles) data. A significantly improved RMS of 3.3 m/s is achieved relative to the single planet fit.}
\label{hd134987bc}
\end{figure}

The AAT and Keck data are shown in Fig.\ref{hd134987} with a single Keplerian curve fit with an orbital period of {\p}~d, a velocity amplitude
of 50.1$\pm$1.5 m/s and an eccentricity of {\e}. The minimum (\msini) mass of the
planet is {\m} \mjup. The RMS to the single Keplerian fit is 6.6~m/s. The activity measure (log~$R_{HK}=$~--5.1) predicts a rotation period of 23---33 d (Wright 2005),  which is
significantly different
from the exoplanet solution. The lack of any observed
chromospheric activity or photometric variations gives us confidence that this solution proposed by Vogt et al. (2000) arises from an exoplanet rather
than from long-period starspots or chromospherically active regions. However, since HD134987 is an inactive star we would expect a substantially lower RMS. Futher to the long-term trend found by Butler et al. (2006a), we
now find a curvature that suggests
a two planet solution with an ~$\sim$5000 d period for the outer planet. Fig. \ref{hd134987bc} shows the best fit double Keplerian to the Keck and AAT data, with an RMS of 3.3 m/s. The
best-fit parameters of the double-Keplerian fit are given in Table \ref{orbit} and are based on treating the offset between the AAT and Keck as a free parameter.

\begin{figure}
\vspace*{1.5cm}
\includegraphics[trim=0 135 40 100,scale=0.5]{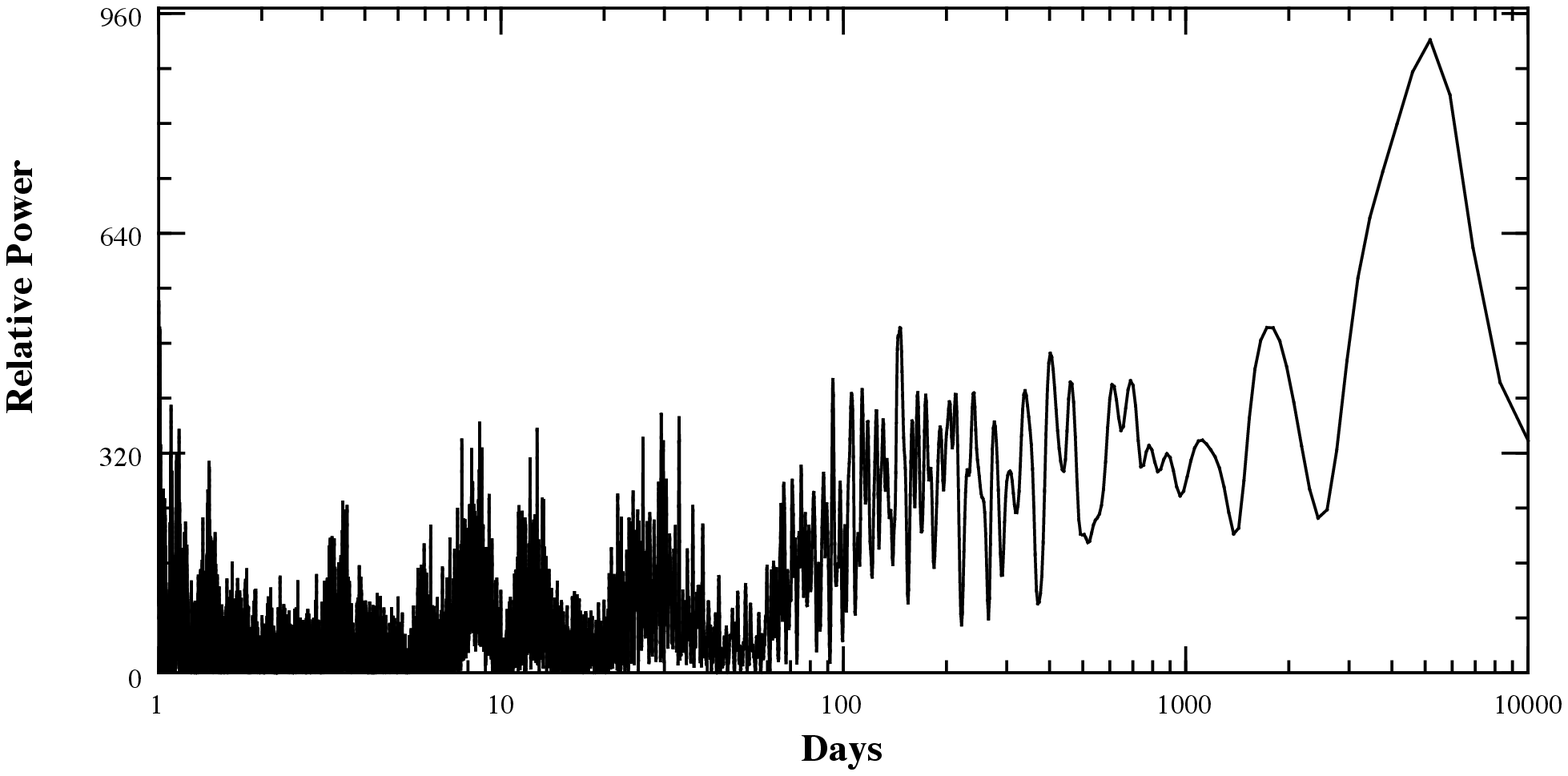}
\vspace*{-4cm}
\newline
\vspace*{-4cm}
\includegraphics[trim=0 135 40 100,scale=0.5]{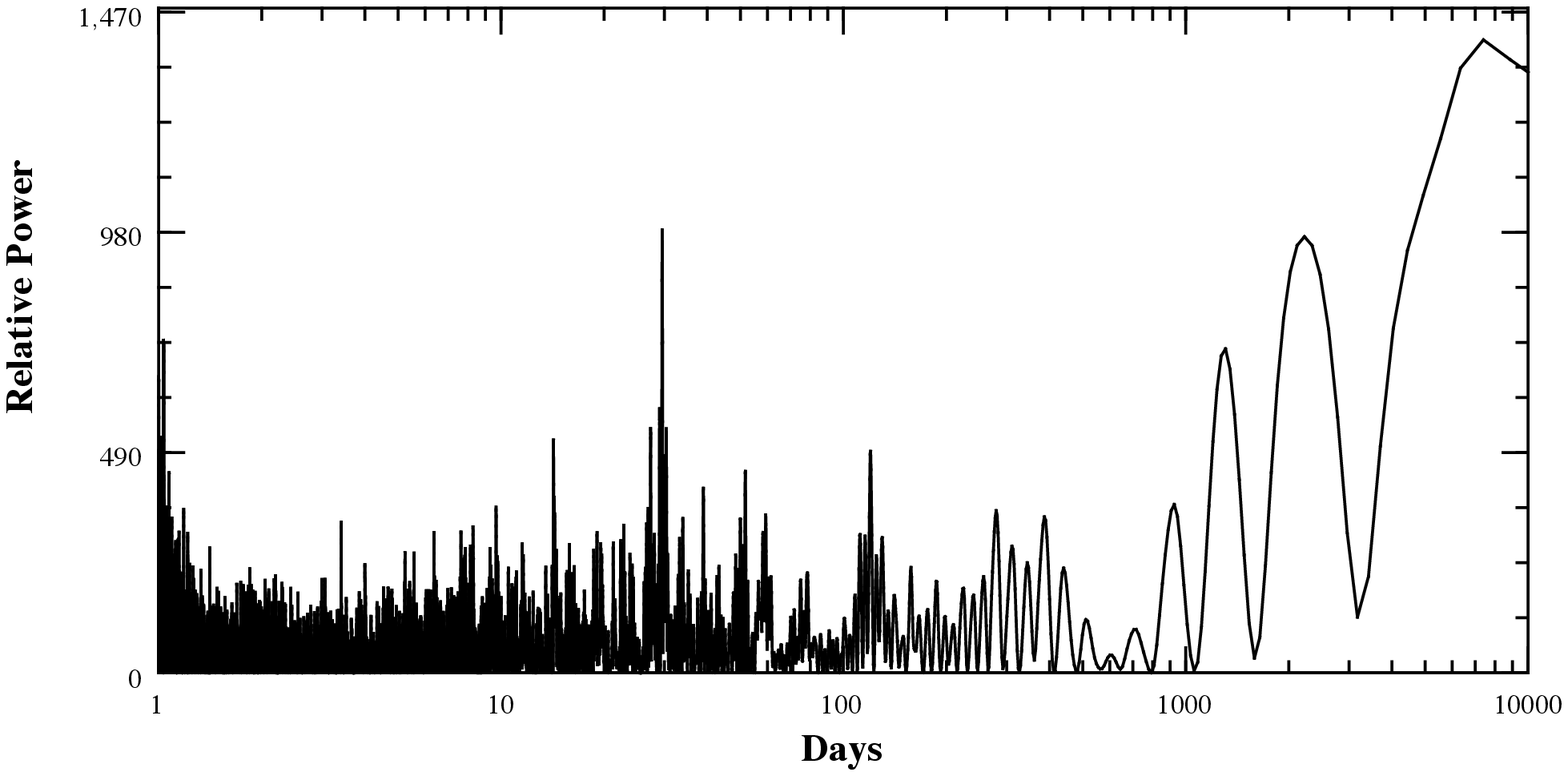}
\newline
\vspace*{-4cm}
\includegraphics[trim=0 135 40 100,scale=0.5]{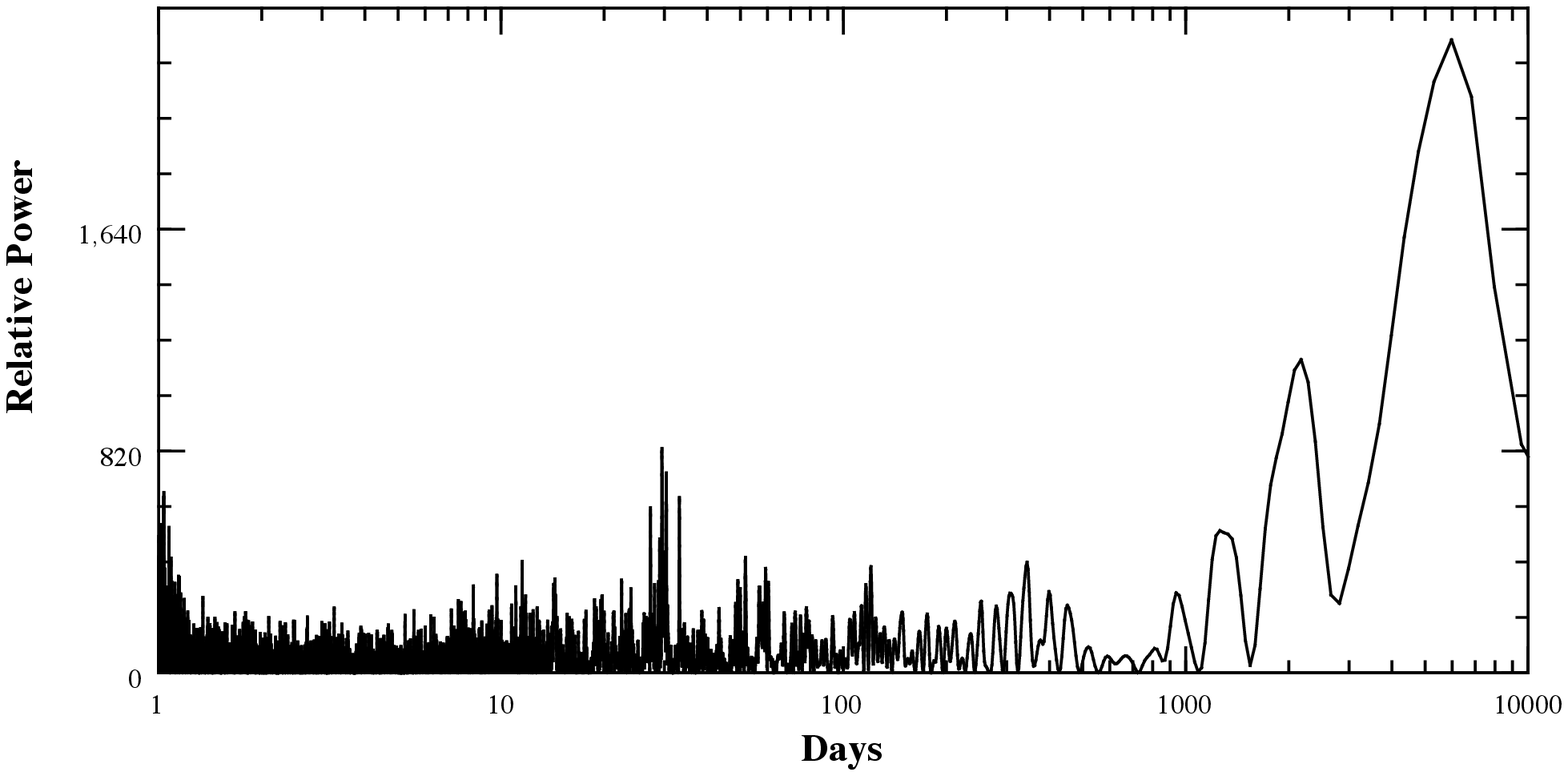}
\caption{\bf{Periodograms of the residuals to a single Keplerian fit to the HD~134987 data: AAT (top), Keck (middle), Combined (bottom). The outputs are taken from the publicly available package {\sc systemic} (Meschiari et al. 2009).}}
\label{power}
\end{figure}

\begin{figure}
\includegraphics[width=120mm,angle=90]{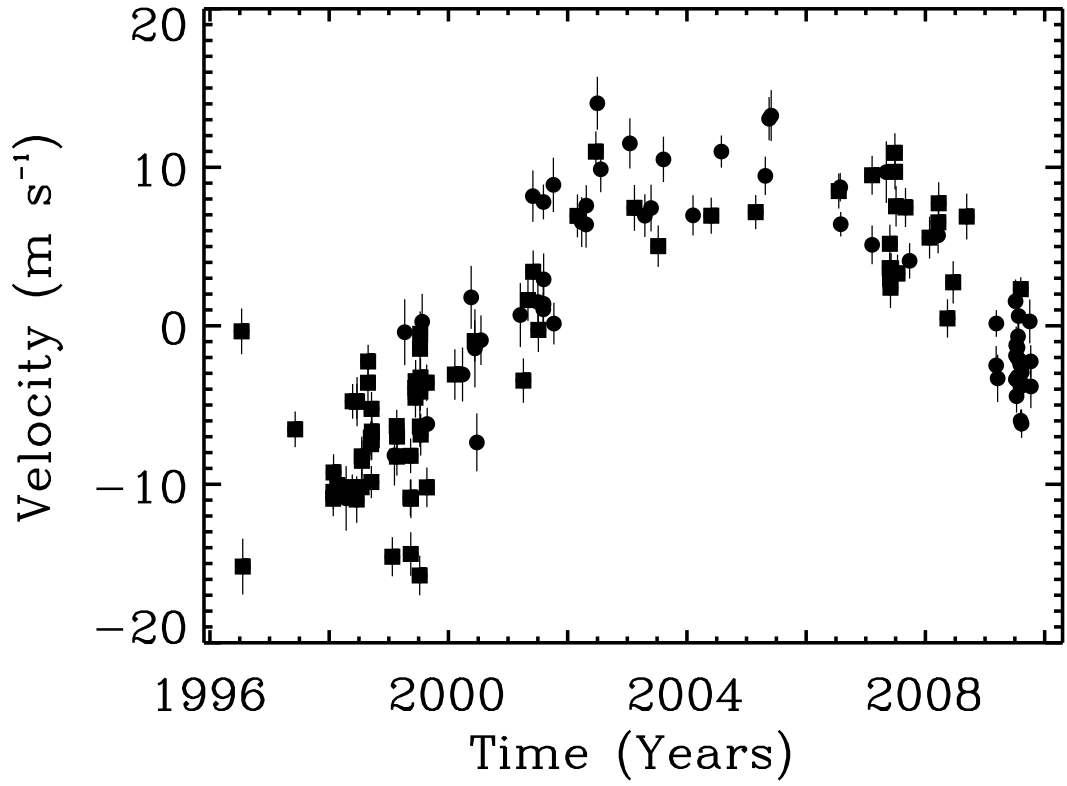}
%\includegraphics[width=110mm,angle=0]{chisq_pvse.ps}
%\includegraphics[width=110mm,angle=0]{chisq_pvsm.ps}
%\vspace*{-4cm}
\caption{Residuals to the 258d Keplerian fit for HD134987b shown in Fig. \ref{hd134987} are shown.}
\label{hd134987c}
\end{figure}

Fig.\ref{power} shows the periodogram of the residuals to HD134987b for the AAT, Keck and combined AAT+Keck datasets. Significant power  is present at periods beyond a few thousand days in the datasets (with low false alarm probabilities computed using {\sc systemic}: AAT -- 5x10$^{-2}$, Keck -- 7x10${^{-5}}$, combined -- 3x10$^{-13}$). We investigated a range of solutions using the AAT and Keck datasets both separately and together. Both datasets produce very similar solutions for the inner planet. For the outer planet we have not definitively seen a full orbital period and thus the orbit is less clearcut. We found that the AAT dataset favours somewhat shorter periods (around
5000~d) and
lower eccentricities  (e$\sim$0.1), while the Keck dataset favours
longer
period solutions with higher eccentricities. The addition of a 5000~d outer planet to either dataset or the combined dataset produces an improvement in RMS by more than a factor of 2 (e.g., Figs \ref{hd134987} and \ref{hd134987bc}). The jitter value which must be used to result in a {\bf reduced} ${\chi_{\nu}^2}$=1 is more than  6~m/s for
a single planet fit, and drops significantly to around 2.6~m/s for a
double planet fit for the combined dataset and 2.3~m/s (AAT) and 3.2~m/s (Keck) for the double planet fit to the separate datasets. These values of jitter are consistent with prediction (2.1~m/s) of
the most recent
activity jitter calibration of J.Wright (priv.comm.). The earliest Keck data have larger uncertainties
than the other Keck observations and appear as outliers in the
residuals plots (Fig. \ref{hd134987c}). We  have therefore checked the sensitivity of the solution
to these three 1996 and 
1997 data points, and find that removal of these data points does not
signiÞcantly change
the best fit orbital parameters and leads to a reduction of only 0.1 m/s in the fit RMS.

\begin{figure}
\includegraphics[width=100mm,angle=90]{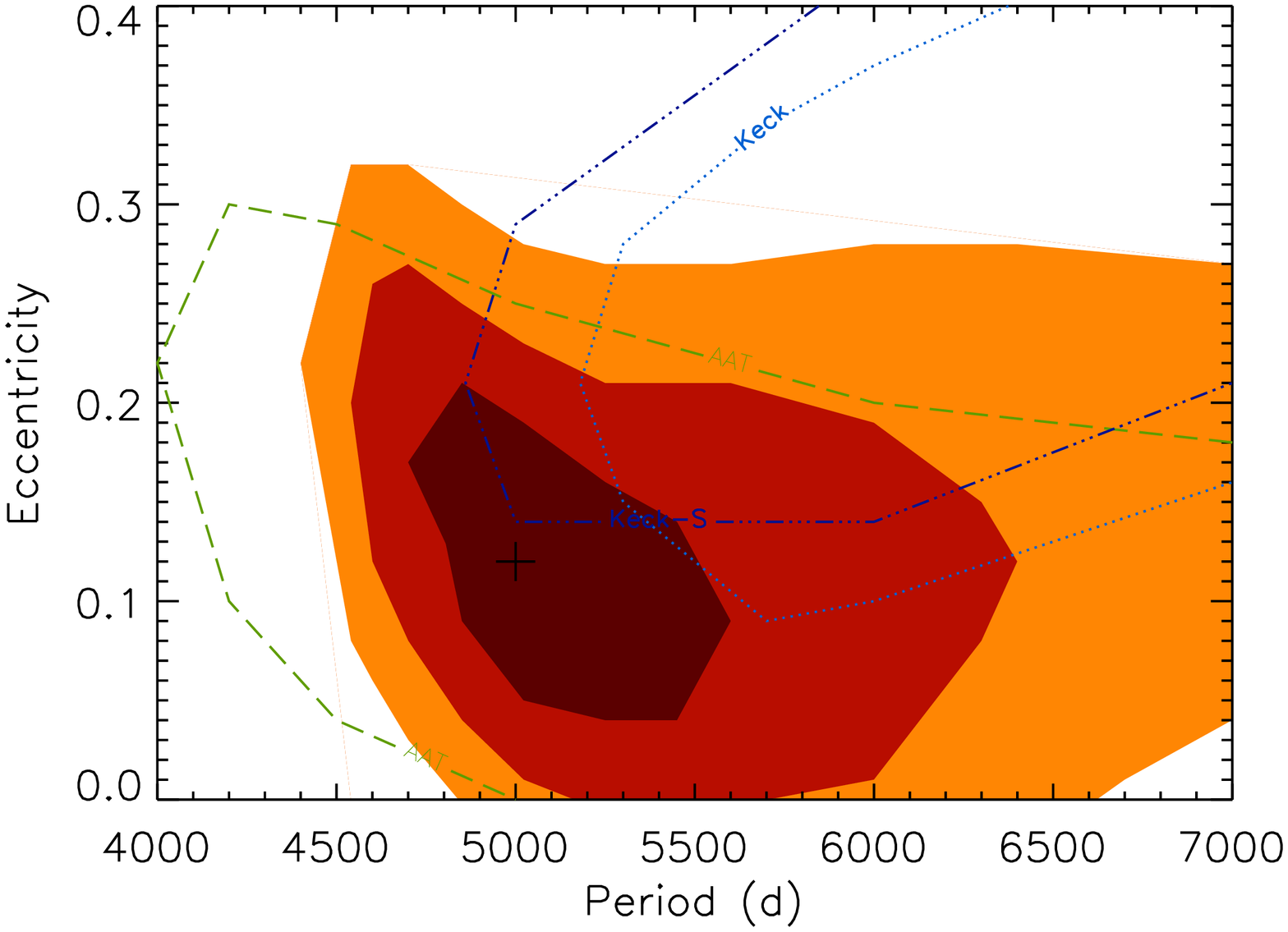}
\includegraphics[width=100mm,angle=90]{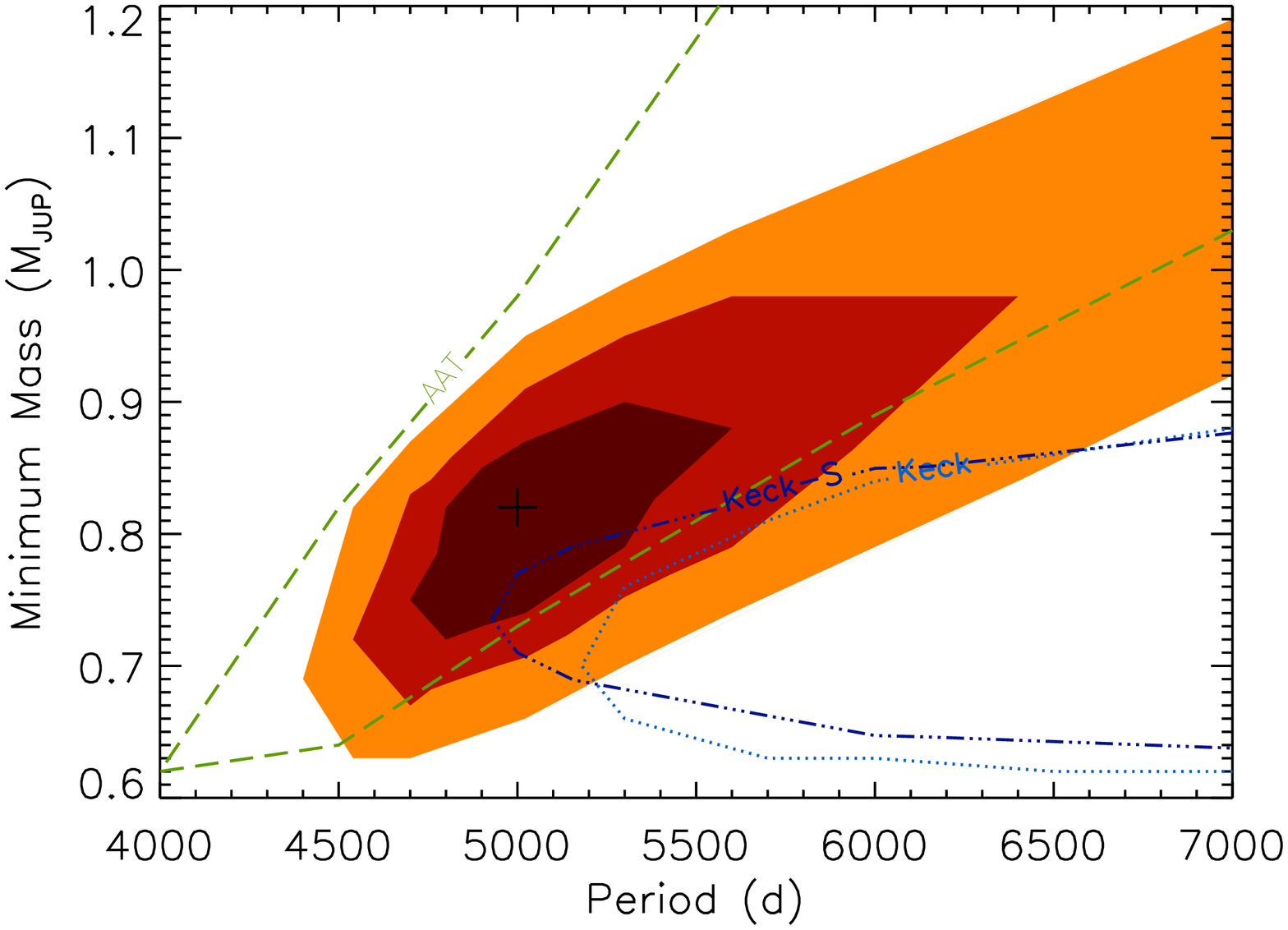}
\caption{The plots show contours of $\chi^2$ for best-fit orbits to the radial velocity data of HD134987c 
in period versus eccentricity (upper plot) and period versus mass (lower plot). The solid brown, red and yellow shading indicate the regions where $\Delta\chi^2$ is up to
2.3, 6.2 and 11.8 from the best-fit minimum $\chi^2$. Dashed green and dotted blue lines show the contours for the AAT and Keck data respectively. For the case of the Keck data a further contour is shown. The dashed-dot-dot-dot contour (dark blue, labelled Keck-S) represents a subset of the Keck data from which velocity values corresponding to 10\% of the highest activity values are removed. The position of the cross marks the best fit solution for the combined dataset.}
\label{new}
\end{figure}

We have not yet seen a full orbital period for HD134987c and so it is difficult to assign a reliable solution for its parameters. In order to better understand how the fit parameters for HD134987c are related, in Fig. \ref{new}, we show contours of best-fit $\chi^2$ for period, mass and eccentricity. The contours indicate best-fit $\chi^2$ solutions increased by 2.3, 6.2 and 11.8, which correspond to 1$\sigma$, 2$\sigma$ and 3$\sigma$ 
confidence levels for systems represented with two degrees of freedom and Gaussian noise. They have been derived allowing all other orbital parameters for HD134987b and c to be best-fit.  These plots highlight the asymmetric nature of the confidence regions due to not having a complete orbital period. These fits are made using the {\sc systemic} package (Meschiari et al. 2009) and assumes no stellar jitter for all data points, however since stellar jitter provides a source of 
pseudo-random noise which may vary on various timescales, e.g., unknown stellar rotation timescale, the noise in the radial velocities
may be non-Gaussian. In addition, to the best fit $\chi^2$ for the joint dataset, 1$\sigma$ contours for the individual AAT (dashed green) and Keck (dotted blue) datasets are shown.

The observing programmes at the AAT and Keck both use the same calibration methodology and follow similar observing and data
reduction strategies. A major difference in their operation for a given star is that Keck achieves
a given S/N in a much shorter exposure time. For HD134987, Keck
integration times range from 27\,s to 10\,min, typically 
1--2 min long, with a median of 84 secs and including 24 measurements of less than a minute. The median of AAT integration times are nearly five times longer (median 400 secs) and with a smaller spread in the range of times from 3.3  to 10\,min.
It is therefore probable that the differences (to the double
planet fit)
seen at each telescope are the result of their sampling astrophysical
noise sources on different time-scales. In order to make the jitter values to achieve a best-fit reduced $\chi^2$ the same for both datasets it
would be necessary
for there to be nearly 2 m/s of additional astrophysical noise (added
in quadrature)
present in the Keck, but not the AAT, data. 

We investigate the importance of the relatively shorter Keck exposure times in a few different ways. We add velocity jitter to all Keck radial velocity errors corresponding to exposure times of less than 200\,sec (the shortest exposure time at the AAT). We do this by adding radial velocity jitter of 4.5$\times$(1-$t_{\rm exp}$/200)~m/s). The scaling factor of 4.5 is chosen so as give resultant best fits requiring the same jitter as the AAT (2.3m/s) to achieve a best-fit  reduced $\chi^2$ of one and is consistent with the higher levels of jitter expected for HD134987c from Wright (2005). However, this procedure as well as removal of the 10\% of the data with the shortest exposure times do not significantly alter the Keck solution.

\begin{figure}
\includegraphics[width=100mm,angle=90]{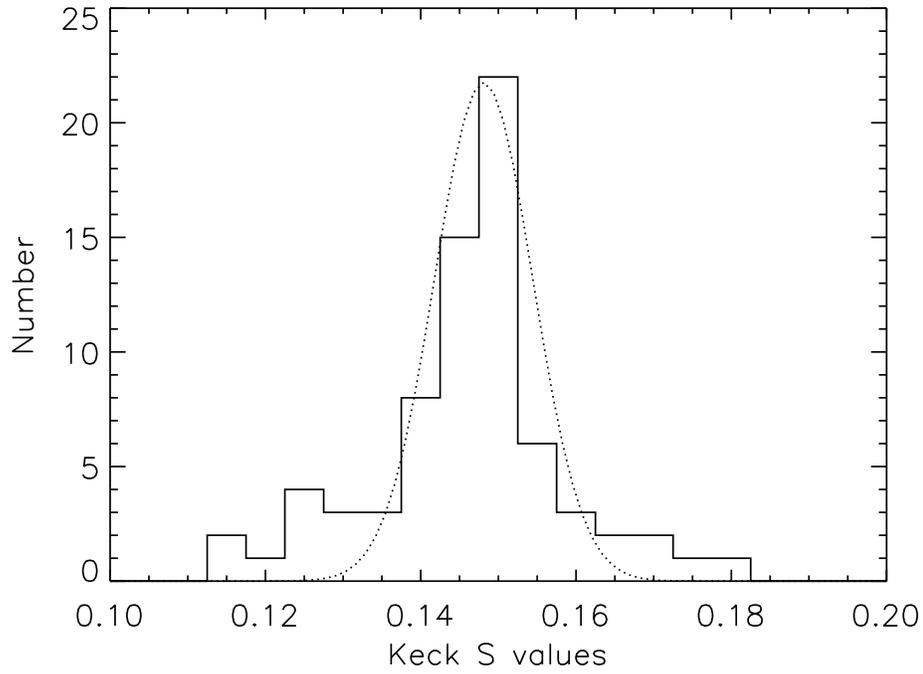}
\caption{Histogram of S values for the Keck dataset plotted with a Gaussian distribution with a full-width half maximum, twice the standard deviation of the S values tabulated in Table \ref{keck_vel}.}
\label{jit}
\end{figure}

The Keck data allows us to gain more direct insight into the importance of the activity of HD134987. The Keck HIRES spectrometer simultaneously covers the the Ca II H\&K lines and the Iodine region. While activity indices such as CaHK do not provide a one-to-one mapping onto stellar jitter and thus can not be used as an input error for the radial velocities, these CaHK lines are the primary method of radial velocity jitter estimation (Wright 2005). From an extraction of this activity measure (e.g., Tinney et al. 2002) the S values for the Keck data are given in Table \ref{keck_vel} and plotted with a Gaussian distribution in Fig. \ref{jit}. This indicates that the distribution of S values is not a particularly good match for a Gaussian. Although the jitter values are the largest source of uncertainty, the assignment of the Gaussian $\sigma$ confidence limits should be robust due the large number of data points available for the fit.
Wright (2005) indicates that in the regime of activity, spectral type and magnitude for $different$ stars in their table 2 for HD134987, there is a factor of 3 difference in radial velocity jitter between 20th and 80th percentile. While this scatter is for $different$ stars we can look at the impact of characterising the jitter radial velocity signal in terms of a linear function varying by a factor of three between the 20th-80th percentiles found by Wright (2005). For the HD134987 Keck data this corresponds to assigning radial velocity jitter values up to 6 m/s. We found that this operation expands the 1-$\sigma$ contours and brings best fit solutions to the Keck dataset to shorter periods. Alternatively, one can obtain a solution with little expansion in the 1-$\sigma$ contours by ignoring radial velocities with high activity values.  Fig. \ref{new} shows 1-$\sigma$ best-fit contours (dashed-dot-dot-dot) for a Keck dataset with the omission of the radial velocity data corresponding to the highest 10\% of S values (that lie in the range 0.164 to 0.182).  The removal of these high S values does not lead to much expansion in the 1-$\sigma$ best-fit contours, removes data points that are spread relatively evenly in time and yields 1-$\sigma$ contours closer to those found for the AAT values.

This result that exposure time has less impact on the solution than the S value can be investigated quantitatively for asteroseismological p-mode jitter. The methodology of O'Toole et al. (2008) allows the p-mode jitter for HD134987 to be derived relatively precisely as a function of exposure time using values from Table \ref{stellarp}. This indicates than the Keck data should present no more than around 0.54 m/s p-mode jitter, compared to a maximum of about 0.36 m/s in AAT data. Since these are significantly smaller than the internal errors, the contributions of other stellar jitter noise sources such as granulation and convection (e.g. Bruntt et al. 2005) are presumably significantly larger. While we are not in a position to quantify the exact noise source responsible for the different jitter values, the low and reasonably consistent values of jitter in agreement with stars of similar spectral type and the improved consistency between datasets on removal of easily identifiable high jitter values indicates that a two planet fit to both AAPS and Keck velocity datasets is consistent with our rudimentary jitter expectations. 
We also note that a periodogram of the Keck S values does not show any significant periodicities nor any peaks with low false alarm probabilities which we would expect if there was a significant spot-induced radial velocity signal in the Keck dataset.

\section{Photometric Observations}

In addition to the spectroscopic observations described and analyzed above, 
we have also acquired high-precision photometric observations of HD~134987 
in seven observing seasons between 1999 March and 2009 June with the T8 
0.80~m automatic photometric telescope (APT), one of seven automatic 
telescopes operated by TSU at Fairborn Observatory in southern Arizona 
(Eaton, Henry \& Fekel 2003).  The APTs can detect short-term, low-amplitude brightness 
variability in solar-type stars resulting from rotational modulation in 
the visibility of active regions, such as starspots and plages e.g., Henry, Fekel \& Hall 1995,
and can also detect longer-term variations 
produced by the growth and decay of individual active regions and the 
occurance of stellar magnetic cycles, e.g., Henry et al. (1995) and Hall et al. 2009. 
The photometric observations can help to establish whether observed radial 
velocity variations are caused by stellar activity or planetary reflex 
motion, e.g., Henry et al. (2000).  Several examples of periodic radial 
velocity variations in solar-type stars caused by photospheric spots and 
plages have been documented by Queloz et al. (2001) and Paulson et al. (2004).  The 
photometric observations are also useful to search for transits of the 
planetary companions e.g., Henry et al. (2000), Sato et al. (2005).

The T8 0.80~m APT is equipped with a two-channel precision photometer 
featuring two EMI 9124QB bi-alkali photomultiplier tubes (PMTs) to make 
simultaneous measurements of a star in Str\"omgren $b$ and $y$ passbands.  
The APT observes each target star (star D) in a quartet with three ostensibly 
constant comparison stars (stars A, B, and C).  From these measurements, 
we compute $b$ and $y$ differential magnitudes for each of the six 
combinations of the four stars: $D-A$, $D-B$, $D-C$, $C-A$, $C-B$, and $B-A$.  
We then correct the Str\"omgren $b$ and $y$ differential magnitudes for 
differential extinction with nightly extinction coefficients and transform 
them to the Str\"omgren photometric system with yearly mean transformation 
coefficients.  Finally, we combine the Str\"omgren $b$ and $y$ differential 
magnitudes into a single $(b+y)/2$ passband to improve the precision of the 
observations. Henry (1999) presents a detailed description of the T8 
automated telescope and photometer, observing techniques, and data reduction 
and quality-control procedures needed for long-term, high-precision 
photometry.

The 419 $D-C$ differential magnitudes of HD~134987 are plotted in the top 
panel of Figure~6.  We chose to analyze the $D-C$ observations for two 
reasons:  (1) they have the smallest standard deviation of the three 
$D-A$, $D-B$, and $D-C$ time series, although not by much (0.00266, 0.00264, 
and 0.00263 mag, respectively), and, more importantly, (2) the $D-C$ 
time series is one year longer than the other two because comparison stars 
$A$ and $B$ had to be replaced after one year due to their variability.  
The three comparison stars $A$, $B$, and $C$ are HD~131992 ($V=6.94$, 
$B-V=0.20$), HD~137076 ($V=8.26$, $B-V=0.41$), and HD~135390 ($V=6.47$, 
$B-V=0.69$).  The standard deviations of the $C-A$, $C-B$, and $B-A$ 
differential magnitudes about their means are 0.00297, 0.00308, and 
0.00266 mag, respectively, comparable to the three standard deviations for 
star $D$ (HD~134987) given above.  These values are somewhat larger than 
our typical precision of ~0.0015 mag because HD~134987 lies at a declination 
of $-25\deg$ and so is observed through high airmass (1.8--2.0).  
The mean precision of the three HD~134987 differential time series is 
0.00264 mag, while the mean precision of the three comparison star time 
series is 0.00290 mag.  Therefore, we have not resolved intrinsic 
brightness variability in HD~134987, and the scatter in the $D-C$ 
measurements can be accounted for by the APT's measurement precision at
high air mass.

Nevertheless, we performed periodogram analyses on all six sets of 
differential magnitudes and find no significant periodicities between 0.03 
and 1000 days.  In particular, a least-squares sine fit to the $D-C$ 
observations phased on companion b's orbital period of 258.187 days gives 
a semi-amplitude of only $0.00035~\pm~0.00017$ mag.  The low level of 
magnetic activity in HD~134987 recorded in Table~1 and the lack of 
detectable photometric variability on the orbital period of companion b 
confirms that the radial velocity variability ($K = 49.5$~m~s$^{-1}$) on 
that period is the result of stellar reflex variability induced by HD~134987b.

\begin{figure}
%\epsscale{0.9}
\includegraphics[width=150mm,angle=0]{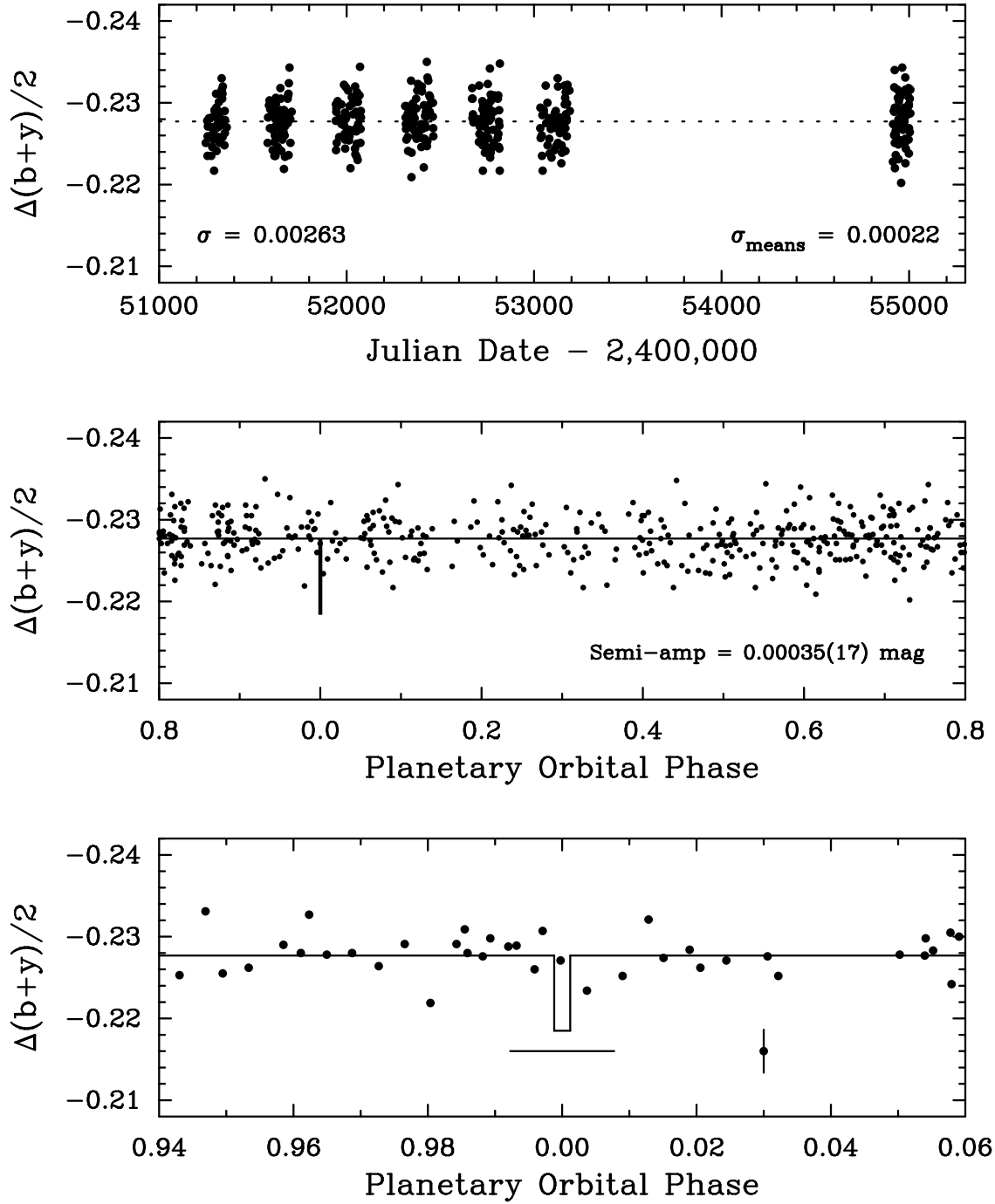}
\caption{Top panel:  the 419 Stro\"mgren $(b+y)/2$ $D-C$ differential 
magnitudes of HD~134987 plotted against heliocentric Julian Date.  The 
standard deviation of the observations from their mean (dotted line) is 
0.00263 mag.  The standard deviation of the yearly means is 0.00022 mag. 
Middle panel:  the observations plotted modulo the 258.187-day orbital 
period of the inner planet.  Phase 0.0 corresponds to the predicted time
of mid transit.  A least-squares sine fit at the orbital period yields
a semiamplitude of only $0.00035\pm0.00017$ mag.  Bottom panel:  the
observations near phase 0.0 plotted on an expanded scale.  The duration
of a central transit is 15 hours while the uncertainty of the transit time
is $\pm2$ days.  Our phase coverage is insufficient to determine whether 
or not companion b transits the star.}
\label{photometry}
\end{figure}

The $D-C$ photometric observations plotted in the top panel of Fig. \ref{photometry}  
cover a range of 3758 days or eleven observing seasons but with a gap of 
four seasons.  So the photometric observations are too few and the orbital 
period of HD~134987c too uncertain ($5000\pm338$) for the photometry to 
determine limits of stellar brightness variability on companion c's orbital 
period.  However, we can look at HD~134987's long-term, year-to-year 
variability for the existing seven observing seasons plotted in Fig. \ref{photometry}.  
The standard deviation of the seven yearly mean magnitudes about their 
grand mean is just 0.000220 mag; the slope of the best-fit line to the 
seven means is $-0.0000245~\pm~0.0000278$~mag yr$^{-1}$.  Interestingly, 
the first six years of our photometry (1999--2004) correspond to the 
interval when the radial velocity residuals to the 258 day period, plotted 
in Figure~4, increased approximately linearly by $\sim22$~m~sec$^{-1}$.  
The best-fit line to those six yearly photometric means has a slope of 
$-0.0000376~\pm~0.0000599$~mag~yr$^{-1}$, which is indistinguishable from 
zero to high precision.  Therefore, the photometric observations also 
provide strong support for the existence of HD~134987c.

With the 419 nightly APT observations of HD~134987, we examine the possibility 
of detecting transits of the inner planet.  The geometric probability for 
transits to occur, given HD~134987b's orbital elements in Table~4, is 
0.65\%, computed from equation (1) of Seagroves et al. (2003).  This is a modest 
improvement over the transit probability (0.52\%) for a circular orbit 
because of the favorable orientation ($\omega=352.8\deg$) of the 
planet's moderately eccentric orbit ($e=0.23$).  The 419 photometric 
measurements in the top panel of Figure~6 are replotted in the middle panel 
phased with the 258.187 day orbital period.  Phase 0.0 corresponds to a 
predicted time of mid transit derived from the orbital elements, 
$T_{transit}=2455027.31$.  The observations near phase 0.0 are replotted 
on an expanded scale in the bottom panel of Fig. \ref{photometry}.  The solid curve in 
the two lower panels approximates the depth (0.85\%) and duration (15 
hours) of a central transit, derived from the orbital elements.  The 
horizontal bar below the predicted transit window in the bottom panel 
represents the $\pm2$ day uncertainty in the time of central transit; the 
vertical error bar to the right of the transit window corresponds to the 
$\pm0.00263$ mag measurement uncertainty of a single observation.  It is 
clear that transits of the inner planet could be detected with the APT, 
but our phase coverage is insufficient to determine whether or not they 
occur.

\section{Discussion}

At a distance of 22~pc, HD~134987 is one of the more nearby exoplanetary systems. The 
combination of the close distance and long period indicates a relatively large angular distance 
from the star of $\gtrsim$0.23 arcsec for an edge-on circular orbit. Such an angular separation will be accessible to the typical 0.2 arcsec
figure of
merit for a current and foreseen high resolution imaging systems on 8m-class telescopes. Only eight other radial-velocity-discovered exoplanets currently exceed a maximum angular separation of 0.2 arcsec  They are $\epsilon$~Eridani~b (Hatzes et al. 2000),  GJ832b (Bailey et al. 2009), 55~Cnc~d (Marcy et al. 2002), HD160691e (McCarthy et al. 2005), GJ849b (Butler et al. 2008b), HD190360b (Naef et al. 2003), 47Uma~c (Fischer et al. 2002), HD154345b (Butler et al. 2006b). Notably, three-quarters of these have been detected using data from the Keck
or AAT.  Despite this promising separation for direct detection, evolutionary
models indicate that
the contrast ratio of HD134987c with its host star will make this a
challenging observation.
Models suggests a  5~Gyr, 2~\mjup~exoplanet at  22~pc will have an H band magnitude of around 35. This will be beyond the reach of even the next generation of high resolution instruments. 
Nevertheless, it should be noted that this estimate is based on
the minimum mass (sin~$i$~=~1) for HD 134987c. Both an inclined orbit,
and a longer period (which is plausible given the
relatively poorly determined period at present) will lead to a larger
mass and improved detectability for HD134987c.
So, imaging observations will be useful to constrain possible orbits and masses of HD~134987c. 

Detection of the astrometric signal from HD~134987~c is more plausible.  The astrometric 
orbit semimajor axis is $\alpha$~sin~$i$ $\gtrsim$ 0.19 mas, which is comparable to the 0.25$\pm$0.06 mas
astrometric orbit determined by Benedict et al. (2002) for GJ 876b. An astrometric 
orbit would enable the inclination to be determined, removing the current sin~$i$ uncertainty 
on the mass. 
%134987 5.8/22*0.81/105/1.047=0.002
%gl832 3.4/4.93*0.64/45/1.047=0.009
%gl876 0.207/4.6*1.89/32/1.047=0.0025

HD134987 joins the family of stars with multiple planets. It appears to be consistent with the broad general properties for multiple planets suggested by Wright et al. (2009). For example, its metallicity of +0.25 is close to that of the mean for exoplanets with long-term trends (+0.20) or multi-planet systems (+0.15, Wright et al. 2009) and the eccentricities of its planets (0.12 and 0.23) are somewhat lower than the 0.25 mean for all exoplanets (excluding tidally circularised ones) and its planetary masses of close to 1\mjup~(the approximate dividing line in mass between higher and lower multiple-planet eccentricities).
From a conservative analysis of exoplanet signals requiring an extra trend to adequately fit the dataset, Wright et al. (2009) suggest that $>$28\% of exoplanets are in multiple systems, a finding likely to be consistent with the AAPS dataset. In order to assess the actual value it will be necessary to assess the detectability of such trends using simulations such as those presented by O'Toole et al. (2009) and Wittenmyer et al. (2009). 

\begin{figure}
\includegraphics[width=100mm,angle=90]{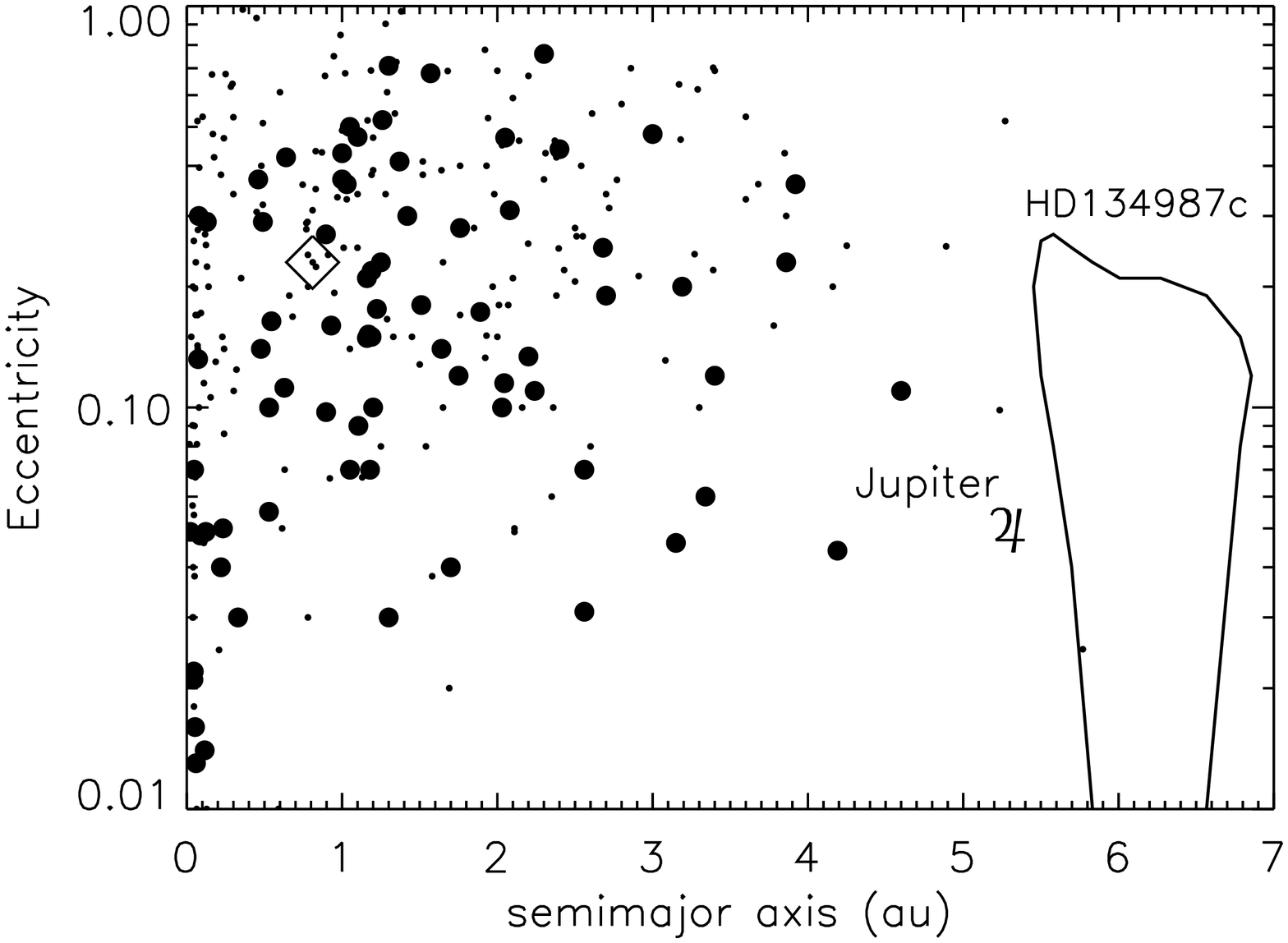}
\includegraphics[width=100mm,angle=90]{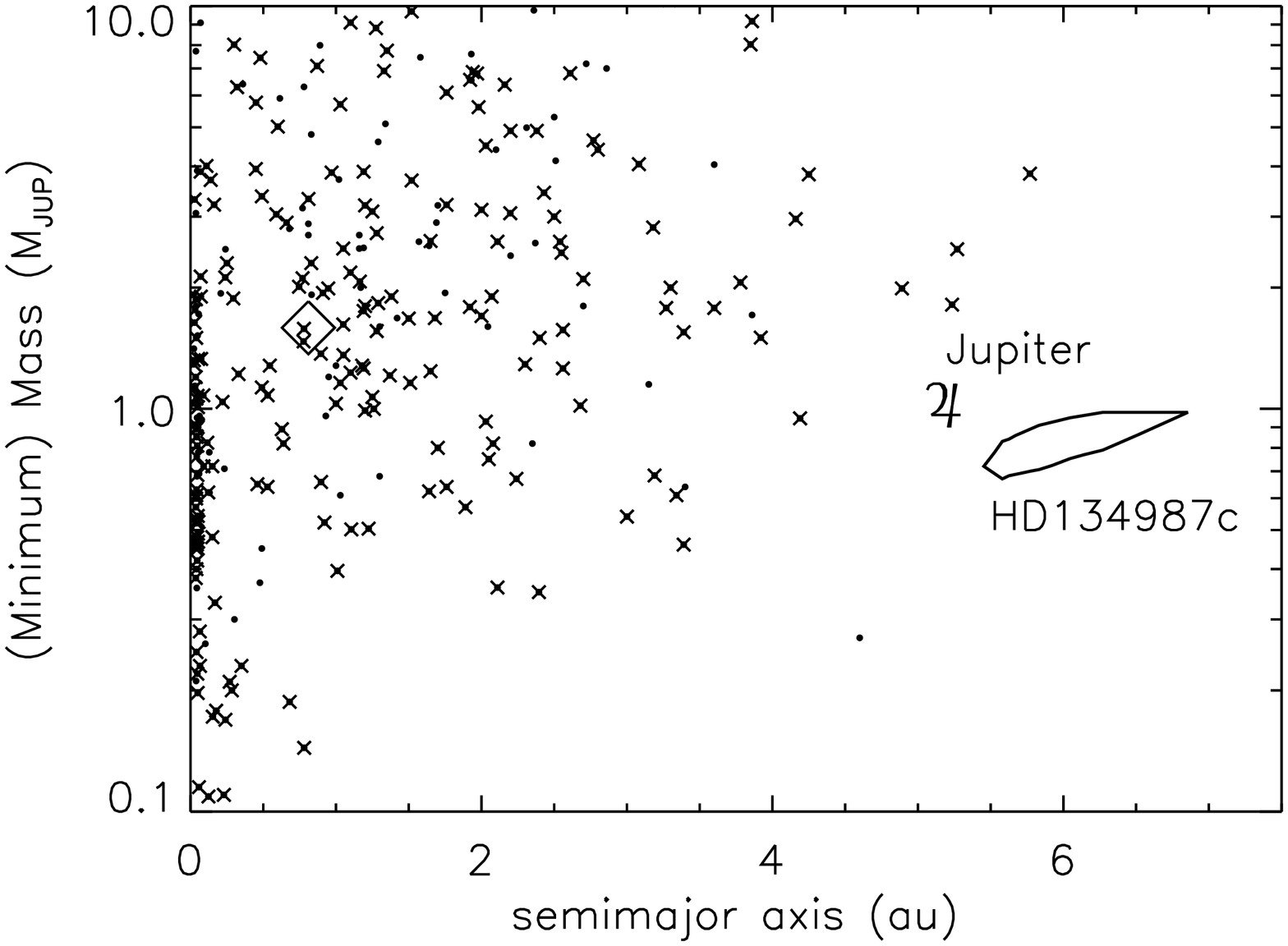}
%\includegraphics[width=110mm,angle=0]{aat_residuals.ps}
%\includegraphics[width=110mm,angle=0]{aat_phased.ps}
%\vspace*{-4cm}
\caption{All exoplanets as recorded at http://exoplanets.eu on 2009 October 27 with planet and star masses, semi-major axes and eccentricity are included as small filled circles. The majority of data in the plots are for radial velocity discovered exoplanets and are included as $M$~sin$i$ values. The upper plot shows semi-major axis as a function of planet mass. Those labelled with crosses have primary masses within 25\% of the Sun. Jupiter's position is marked by the traditional mythological symbol. Jupiter's semi-major axis and eccentricity are plotted as 5.20au and 0.0489 respectively from the Astronomical Almanac (2000). The best fit orbit for HD134987b is indicated by diamonds with the 2-$\sigma$ best-fit contour for HD134987c orbit. In the lower plot points plotted as large filled circles are those whose ratios of planet mass to star mass fall within 50\% of that of Jupiter and the Sun.}  
\label{au_mp2009}
\end{figure}

The fit to HD134987c indicates a relatively small semi-amplitude ($\sim$10m/s) in comparison to most of the other long-period exoplanets announced to date. Fig. \ref{au_mp2009} indicates that the orbital solution for HD134987c is rather more reminiscent of Jupiter than other exoplanets discovered to date.  With a longer semi-major axis than Jupiter and similar eccentricity, the discovery of HD134987c signals that we have sensitivity to radial velocity planets with Jupiter-like periods around Sun-like stars. As the second decade of data is now being gathered by the AAT and
Keck Telescopes, we can be confident that our long-term precision is
sufficient to empirically constrain the incidence of 
exoplanets with Jupiter-like periods around Sun-like stars. This will enable us to ascertain just how common our Solar System might be and be able to make comparison with developing theoretical predictions (e.g., Mordasini et al. 2009). As our temporal baseline 
extends, we will become sensitive to longer-period planets, e.g., a true Saturn 
analog would require 15 more years of observation to fully sense. As our precision improves we will become sensitive to lower-mass longer-period exoplanets, which migration scenarios for planets around solar-type stars suggest is a rich domain (e.g., Schlaufman, Lin \& Ida 2009). 

\section*{Acknowledgments}
We thank the Anglo-Australian and Keck time assignment
committees for continuing allocations of telescope time.  We are grateful
for the extraordinary support we have received from the
AAT technical staff  -- E. Penny, R. Paterson, D. Stafford,
F. Freeman, S. Lee, J. Pogson, S. James, J. Stevenson, K. Fiegert and W. Campbell.  
We gratefully
acknowledge the UK and Australian government support of the
Anglo-Australian Telescope through their PPARC, STFC and DETYA funding
(HRAJ, CGT); NSF grant AST-9988087, NASA grant
NAG5-12182, STFC grant PP/C000552/1, ARC Grant DP0774000
and travel support from the Carnegie Institution
of Washington (to RPB) and from the Anglo-Australian Observatory (to CGT, BDC
and JB).  SSV gratefully acknowledges support from NSF grant AST-0307493. GWH acknowledges support from NASA, NSF, Tennessee State University, and
the state of Tennessee through its Centers of Excellence program. This research has made use of the SIMBAD and exoplanet.eu 
databases, operated at CDS, Strasbourg and Paris Observatory respectively.  The referee is thanked for making suggestions which led to substantial  improvements in the paper.

\newpage

\begin{deluxetable}{lrrrrr}
%\begin{table}
% \centering
% \begin{minipage}{140mm}
\tablecaption{The stellar parameters for HD~134987 are tabulated below.}
\tablewidth{0pt}
%  \begin{tabular}{@{}lrrrrr@{}}
%Parameter &  H2007 & T2008& S2008&T2008 \\
\tablehead{
Parameter &  Value &Reference}
\startdata
Spectral Type & G5V &  Cenarro et al. (2007)\\
log $R'_{\rm HK}$ & -5.04 & Jenkins et al. (2006) \\
& -5.13 & Saffe et al. (2005) \\
%Hipparcos $N_{\rm obs}$ & 113 & Leuwen (2007)\\
Variability( $\sigma$) & 0.0013 & van Leuwen (2007)\\
Distance(pc) & 22.2$\pm$1.1 & van Leuwen (2007)\\
log (L$_{\rm star}/$L$_\odot$) &  1.80$\pm$0.14 & van Belle \& von Braun (2009) \\
&  1.43$\pm$0.02 & Sousa et al. (2008) \\
R$_{\rm star}$/R$_\odot$ & 1.25$\pm{0.04}$ & van Belle \& von Braun (2009) \\
M$_{\rm star}$/M$_\odot$ & 1.07$\pm^{0.03}_{0.08}$ & Holmberg et al. (2007) \\
 & 1.05$\pm^{0.07}_{0.05}$& Takeda et al. (2007a)\\
&  1.10$\pm^{0.07}_{0.04}$ & Takeda (2007b)\\
T$_{\rm eff}$ (K)&    5636 &Holmberg et al. (2007) \\
&    5766 & Takeda et al. (2007b)\\
&    5740$\pm$23& Sousa et al. (2008)\\
&    5585$\pm$50 &    van Belle \& von Braun (2009)\\
& 5623 $\pm$57 & Bond et al. (2006)\\
$[$Fe/H$]$ & 0.20     &  Holmberg et al. (2007)\\
& 0.28 & Takeda et al. (2007b)\\
& 0.25$\pm$0.02& Sousa et al. (2008)\\
U,V,W (km~s$^{-1}$) &-21,-41,20 &  Holmberg et al. (2007) \\
 & -9.8,-25.4,28.4& Takeda et al. (2007b)\\
Age (Gyr) & 8.4$\pm^{1.6}_{1.4}$  & Takeda et al. (2007a)\\
& 11.1$\pm^{1.5}_{3.7}$ & Saffe et al. (2008)\\
jitter  (m~s$^{-1}$)  &  3.5              &       Butler et al. (2006a)\\
$v$~sin~$i$ (km/s)  &  2.17    &     Butler et al. (2006a)\\
%distance         &   45.0$\pm$2.3           &         36.0$\pm$1.1     \\
%Hipparcos $\pi$ & 22.2$\pm$1.1 & 27.8$\pm$0.9\\
\label{stellarp}
%\end{tabular}
%\end{minipage} 
%\end{table}
\enddata
\end{deluxetable}

\begin{deluxetable}{rrr}
\tablecaption{Relative radial velocities (RV) and error in m/s are given for the AAT dataset. Julian
Dates (JD) are heliocentric.
RVs are barycentric but have an arbitrary
zero-point determined by the radial velocity of the template.}
\tablewidth{0pt}
\tablehead{
JD$~~$ & RV$~~$ & Unc.$~~$ \\
(-2450000)   &  (m\,s$^{-1}$) & (m\,s$^{-1}$)
}
\startdata
%\tableline
     917.2282  &   -41.8  &  2.0 \\
 1213.2775  &   -61.6  &  1.9 \\
 1276.0475  &   -50.4  &  2.1 \\
 1382.9573  &    30.1  &  1.8 \\
 1413.8813  &   -16.4  &  1.0 \\
 1630.2677  &    36.1  &  1.7 \\
 1683.0609  &   -20.8  &  2.0 \\
 1706.0960  &   -43.3  &  2.5 \\
 1717.9564  &   -56.0  &  1.8 \\
 1742.9340  &   -57.7  &  1.6 \\
 1984.2154  &   -51.4  &  2.0 \\
 2060.9717  &   -35.8  &  1.6 \\
 2091.9394  &   -11.9  &  1.3 \\
 2124.8927  &    33.8  &  1.0 \\
 2125.0410  &    34.2  &  1.3 \\
 2125.8890  &    41.6  &  1.1 \\
 2125.9847  &    35.2  &  1.2 \\
 2126.9191  &    37.7  &  1.6 \\
 2186.8784  &     0.6  &  1.7 \\
 2189.8632  &   -11.8  &  1.3 \\
 2360.2379  &     7.4  &  1.6 \\
 2387.1061  &    42.9  &  1.5 \\
 2388.1532  &    44.9  &  1.3 \\
 2455.9877  &    -6.8  &  1.7 \\
 2476.9724  &   -29.5  &  1.4 \\
 2655.2513  &    52.5  &  1.6 \\
 2747.1502  &   -39.9  &  1.3 \\
 2785.0880  &   -50.4  &  1.5 \\
 2860.8964  &   -10.0  &  1.4 \\
 3042.2517  &   -50.8  &  1.3 \\
 3215.9389  &     7.3  &  1.0 \\
 3485.0592  &    -7.4  &  1.2 \\
 3508.1736  &   -25.2  &  1.4 \\
 3521.0784  &   -33.2  &  1.6 \\
 3943.9202  &    49.4  &  0.9 \\
 3946.9329  &    47.4  &  0.8 \\
 4139.2578  &   -28.8  &  1.2 \\
 4226.0985  &    36.0  &  1.9 \\
 4368.8908  &   -48.2  &  1.1 \\
 4543.2849  &   -34.2  &  1.1 \\
 4899.2183  &   -47.7  &  1.2 \\
 4900.2294  &   -44.4  &  0.8 \\
 4908.2357  &   -42.0  &  1.5 \\
 5017.9707  &     4.8  &  1.4 \\
 5020.0140  &    -2.8  &  1.1 \\
 5020.9416  &    -2.5  &  1.0 \\
 5021.9442  &    -3.2  &  1.2 \\
 5023.9401  &    -9.0  &  1.0 \\
 5029.9292  &   -13.3  &  1.1 \\
 5030.8891  &   -15.1  &  0.9 \\
 5032.0031  &   -17.6  &  0.9 \\
 5032.9382  &   -16.1  &  1.0 \\
 5036.9205  &   -19.3  &  1.2 \\
 5044.9871  &   -31.4  &  1.2 \\
 5046.0131  &   -31.4  &  0.7 \\
 5046.9415  &   -32.0  &  0.9 \\
 5047.8949  &   -34.5  &  0.8 \\
 5048.9788  &   -37.6  &  0.7 \\
 5054.8813  &   -42.6  &  0.9 \\
 5055.9357  &   -40.2  &  0.5 \\
 5104.8746  &   -57.2  &  1.4 \\
 5110.8763  &   -60.2  &  1.0 \\
 5111.8779  &   -61.8  &  1.4 \\
\enddata
\label{aat_vel}
\end{deluxetable}

\begin{deluxetable}{rrrrr}
\tablecaption{Relative radial velocities, errors in m/s, S values and exposure times ($t_{\rm exp}$) are given for the Keck dataset. Julian
Dates are heliocentric.
RVs are barycentric but have an arbitrary
zero-point determined by the radial velocity of the template.}
\tablewidth{0pt}
\tablehead{
JD$~~$ & RV$~~$ & Unc.$~~$ & S$~~$ & $t_{\rm exp}$\\
(-2450000)   &  (m\,s$^{-1}$) & (m\,s$^{-1}$)&  & s 
}
\startdata
  276.8020  &    -9.3  &  1.4  &  0.122  &600\\
  283.8984  &   -15.4  &  1.8  & -0.020  &300\\
  604.8935  &    41.2  &  1.1  &  0.135  &150\\
  838.1755  &    40.0  &  1.1  &  0.151  &143\\
  839.1727  &    41.2  &  1.1  &  0.150  &143\\
  840.1707  &    43.1  &  1.2  &  0.156  &143\\
  863.1203  &    37.6  &  1.2  &  0.148  & 80\\
  954.9176  &   -49.3  &  0.8  &  0.143  &143\\
  956.9547  &   -44.6  &  1.1  &  0.153  &143\\
  981.8126  &   -54.8  &  1.5  &  0.151  &120\\
  982.8190  &   -48.6  &  1.1  &  0.154  &140\\
  983.8498  &   -48.6  &  1.6  &  0.154  &110\\
 1011.8011  &   -48.6  &  1.1  &  0.150  & 60\\
 1012.8005  &   -46.2  &  1.2  &  0.138  &100\\
 1013.8006  &   -46.1  &  1.3  &  0.128  & 80\\
 1050.7730  &   -13.2  &  1.1  &  0.170  & 80\\
 1051.7547  &   -10.7  &  1.1  &  0.152  &268\\
 1068.7306  &     7.1  &  1.0  &  0.174  &100\\
 1069.7193  &     8.2  &  1.0  &  0.172  &100\\
 1070.7242  &     9.9  &  1.0  &  0.164  &110\\
 1071.7229  &     8.8  &  1.0  &  0.182  &120\\
 1072.7204  &    14.9  &  1.1  &  0.161  & 80\\
 1073.7196  &    14.4  &  1.2  &  0.051  &120\\
 1074.7070  &    16.4  &  1.0  &  0.166  &120\\
 1200.1581  &   -48.3  &  1.2  &  0.147  &100\\
 1227.0883  &   -49.0  &  1.0  &  0.148  &120\\
 1228.1038  &   -51.1  &  1.2  &  0.157  &120\\
 1229.1161  &   -49.9  &  1.1  &  0.156  &140\\
 1310.8892  &   -18.2  &  1.1  &  0.143  &100\\
 1311.9101  &   -14.3  &  1.1  &  0.144  &100\\
 1312.9239  &   -19.3  &  1.4  &  0.138  & 70\\
 1314.0005  &   -14.5  &  1.2  &  0.150  &100\\
 1340.8393  &    30.9  &  1.1  &  0.148  &119\\
 1341.8853  &    31.8  &  1.3  &  0.154  &119\\
 1342.8787  &    34.2  &  1.3  &  0.150  & 60\\
 1367.7877  &    39.2  &  1.3  &  0.150  & 90\\
 1368.7558  &    48.3  &  1.3  &  0.146  & 55\\
 1369.7821  &    52.9  &  1.3  &  0.156  & 60\\
 1370.8677  &    53.4  &  1.4  &  0.155  & 90\\
 1371.7599  &    49.4  &  1.2  &  0.149  & 50\\
 1372.7678  &    49.7  &  1.2  &  0.145  & 60\\
 1373.7712  &    45.5  &  1.3  &  0.151  & 40\\
 1410.7258  &     4.3  &  1.1  &  0.153  &119\\
 1411.7251  &    -3.6  &  1.3  &  0.143  &119\\
 1583.1620  &     8.2  &  1.6  &  0.153  & 70\\
 1704.8398  &   -27.8  &  1.3  &  0.148  & 57\\
 2002.9877  &   -46.3  &  1.4  &  0.128  &213\\
 2030.9596  &   -40.7  &  1.3  &  0.134  &194\\
 2062.8470  &   -25.1  &  1.3  &  0.119  & 68\\
 2094.7919  &     4.4  &  1.4  &  0.119  & 87\\
 2334.1494  &   -10.6  &  1.4  &  0.127  & 56\\
 2446.9002  &    14.6  &  1.3  &  0.135  & 74\\
 2683.1724  &    40.0  &  1.5  &  0.140  & 50\\
 2828.8530  &   -28.8  &  1.3  &  0.130  & 52\\
 3153.8956  &    49.5  &  1.1  &  0.135  & 42\\
 3426.0749  &    61.6  &  1.1  &  0.142  &446\\
 3934.7661  &    57.9  &  1.1  &  0.144  & 59\\
 4139.1543  &   -10.3  &  1.2  &  0.149  & 40\\
 4246.9373  &    16.4  &  0.9  &  0.155  &173\\
 4247.9507  &    16.7  &  1.2  &  0.154  & 56\\
 4248.9033  &    13.2  &  1.0  &  0.153  & 27\\
 4251.8409  &     8.9  &  1.3  &  0.152  & 41\\
 4255.8205  &     5.3  &  1.0  &  0.151  & 27\\
 4278.7856  &   -10.1  &  1.2  &  0.153  & 42\\
 4279.7868  &   -12.1  &  1.1  &  0.150  & 49\\
 4285.7953  &   -18.7  &  1.3  &  0.161  & 93\\
 4294.8522  &   -28.6  &  1.2  &  0.151  & 53\\
 4343.7227  &   -36.2  &  1.2  &  0.159  & 42\\
 4491.1711  &    37.2  &  1.3  &  0.149  & 55\\
 4545.0821  &   -20.5  &  1.3  &  0.150  & 36\\
 4547.0645  &   -20.6  &  1.3  &  0.149  & 36\\
 4600.9843  &   -43.3  &  1.2  &  0.148  & 84\\
 4635.8703  &   -31.3  &  1.3  &  0.152  & 27\\
 4718.7306  &    61.8  &  1.4  &  0.159  & 50\\
 5049.8434  &   -15.8  &  0.8  &  0.177  & 59\\
\enddata
\label{keck_vel}
\end{deluxetable}

\begin{deluxetable}{lrrr}
\tablecaption{Orbital parameters with standard errors for double Keplerian fit to HD~134987 dataset based on our best fit solution.}
\tablewidth{0pt}
\tablehead{
%\begin{table}
%\centering
%\begin{minipage}{180mm}
%\caption{Orbital parameters fit to our HD~134987 dataset.}
%\begin{tabular}{@{}lrrr@{}}
%\hline
& HD~134987b &  HD~134987c} 
\startdata 
Orbital period $P$ (d) &  {\p} & {\ppe}  \\
Velocity amplitude $K$ (m~s$^{-1}$) &{\sa} & {\sasae}\\
Eccentricity $e$  &{\e} & {\eee}  \\
$\omega$ (deg)    & {\om} & {\omome} \\
Periastron Time (JD)  &{\pt} & {\ptpte} \\
$M$sin$i$ (\mjup)     &{\m}       & {\mme}\\
a (au)               & {\au} & {\auaue}  \\
`Zero point RV Offset'  & {\off} \\
`AAT Offset & {\aoff} \\
\label{orbit}
%\end{tabular}
%\end{minipage}
%\end{table}
\enddata
\end{deluxetable}

\newpage

%\newpage

\newpage

\newpage

\clearpage
\end{document}